\documentstyle[mprocl]{article}

\input{psfig.sty}
\def\ie{{\it i.e.}}
\def\eg{{\it e.g.}}

\def\epem{\ifmmode e^+e^-\else $e^+e^-$\fi}
\def\mpl{\ifmmode \overline M_{Pl}\else $\bar M_{Pl}$\fi}
\def\be{\begin{equation}}
\def\ee{\end{equation}}
\def\bea{\begin{eqnarray}}
\def\eea{\end{eqnarray}}

\begin{document}

\rightline{\vbox{\halign{&#\hfil\cr
&SLAC-PUB-8339\cr
&January 2000\cr}}}
\vspace{0.8in}

\title{{TEV GRAVITY AND KALUZA-KLEIN EXCITATIONS IN $e^+e^-$ AND $e^-e^-$ 
COLLISIONS}
\footnote{To appear in the {\it Proceedings of the $3^{nd}$ International 
Workshop on $e^-e^-$ Interactions at TeV Energies}, Santa Cruz, CA, 
10-12 December 1999}
}

\author{ {T.G. RIZZO}
\footnote{Work supported by the Department of Energy, 
Contract DE-AC03-76SF00515}
}

\address{Stanford Linear Accelerator Center,\\
Stanford University, Stanford, CA 94309, USA\\
E-mail: rizzo@slacvx.slac.stanford.edu}

\maketitle\abstracts{We review the capability of $e^+e^-$ and $e^-e^-$ 
colliders to detect the virtual exchange of Kaluza-Klein towers of gravitons 
within the large extra dimension scenario of Arkani-Hamed, Dimopoulos and 
Dvali and in the localized gravity model of Randall and Sundrum with 
non-factorizable geometry.}

\section{Introduction}

The exciting possibility that the gravity may become strong at the TeV scale 
due to the existence of extra dimensions 
offers some hope of solving the hierarchy problem. In this talk we review 
some of the experimental signatures for two such theories at $e^\pm e^-$ 
colliders. As we will see the predictions of these models are quite distinct 
but, in either case, lepton colliders will play a significant role in probing 
their detailed structure.

\section{The ADD Model}

In the model of Arkani-Hamed, Dimopoulos and Dvali(ADD)~{\cite {nima}}, gravity 
is allowed to live in $n$ `large' extra dimensions, \ie, `the bulk', while the 
Standard Model(SM) fields lie on a 3-D surface or brane, `the wall'. Gravity 
then becomes strong in the full $4+n$-dimensional space at a scale $M_s\sim$ a 
few TeV which is far below the conventional Planck scale, $M_{pl}\sim 10^{19}$ 
GeV. The scales $M_s$ and $M_{pl}$ are simply related via Gauss' Law:
$M_{pl}^2=V_nM_s^{n+2}$, with $V_n$ being the volume of the compactified 
extra dimensions. For $n$ extra dimensions of the same size, $V_n\sim R^n$ and 
one finds that $R \sim 10^{30/n-19}$ meters assuming $M_s\sim 1$ TeV. Note that 
for separations between two masses less than $R$ the gravitational force law 
becomes $1/r^{2+n}$. For $n=1$, $R\sim 10^{11}$ meters and is thus excluded, 
but, for $n=2$ one obtains $R \sim 0.1$~mm, which is at the edge of the 
sensitivity for existing experiments~{\cite {test}}. 
Astrophysics~{\cite {astro}} requires 
that $M_s>110$ TeV for $n=2$ but only $\geq $ a few TeV for $n>2$. 
The Feynman rules for this scenario can be found in Ref.~{\cite {pheno1}}. 
Upon compactification, one 
finds that all of the members of the Kaluza-Klein(KK) tower of gravitons 
couples exactly as does the zero mode.

Outside of table top experiments that probe Newtonian gravity at short 
distances and astrophysics, the two ways of probing 
this scenario at colliders are 
via the emission of KK towers of gravitons in scattering processes or through 
the exchange of KK graviton towers between SM fields~{\cite {pheno1}}, 
with which we will be interested here. The virtual 
exchange of graviton towers either leads to modifications in SM cross 
sections and asymmetries or to new processes not allowed in the SM at the tree 
level. In the case of exchange the amplitude is proportional to the sum over 
the propagators of the entire KK tower which naively diverges when $n>1$. 
This summation can 
either be regulated by a brute force cut-off, by the tension of the 
3-brane~{\cite {wow}}, or through the finite extent of the SM fermion wave 
functions in the additional dimensions~{\cite {schm}}. The differential cross 
sections then become relatively $n$ insensitive functions of the effective 
cut-off scale, which we will here call $M_H$~{\cite {funny}}, and the overall 
sign of the dimension-8 operator induced by the KK tower, $\lambda$. We expect 
that $M_H$ and the scale $M_s$ are qualitatively similar, being exact equal in
the case of a very stiff brane. In this 
virtual exchange, all of the gravitons act coherently and, due to their 
relatively tiny mass splittings, sum to a result which is only $M_H\sim 1$ TeV 
suppressed instead of Planck mass suppressed. Individual resonances associated 
with a given graviton exchange are smeared out and are not observable. 
A characteristic feature in all cases is 
the rapid growth with energy of the graviton contribution to the amplitude; 
relative to the pure SM, interference terms go as $\sim s^2/M_H^4$ whereas 
the pure gravity terms behave as $\sim s^4/M_H^8$. Thus we expect these KK 
contributions to cross section amplitudes to turn on rather rapidly.

\begin{figure}[htbp]
\centerline{
\psfig{figure=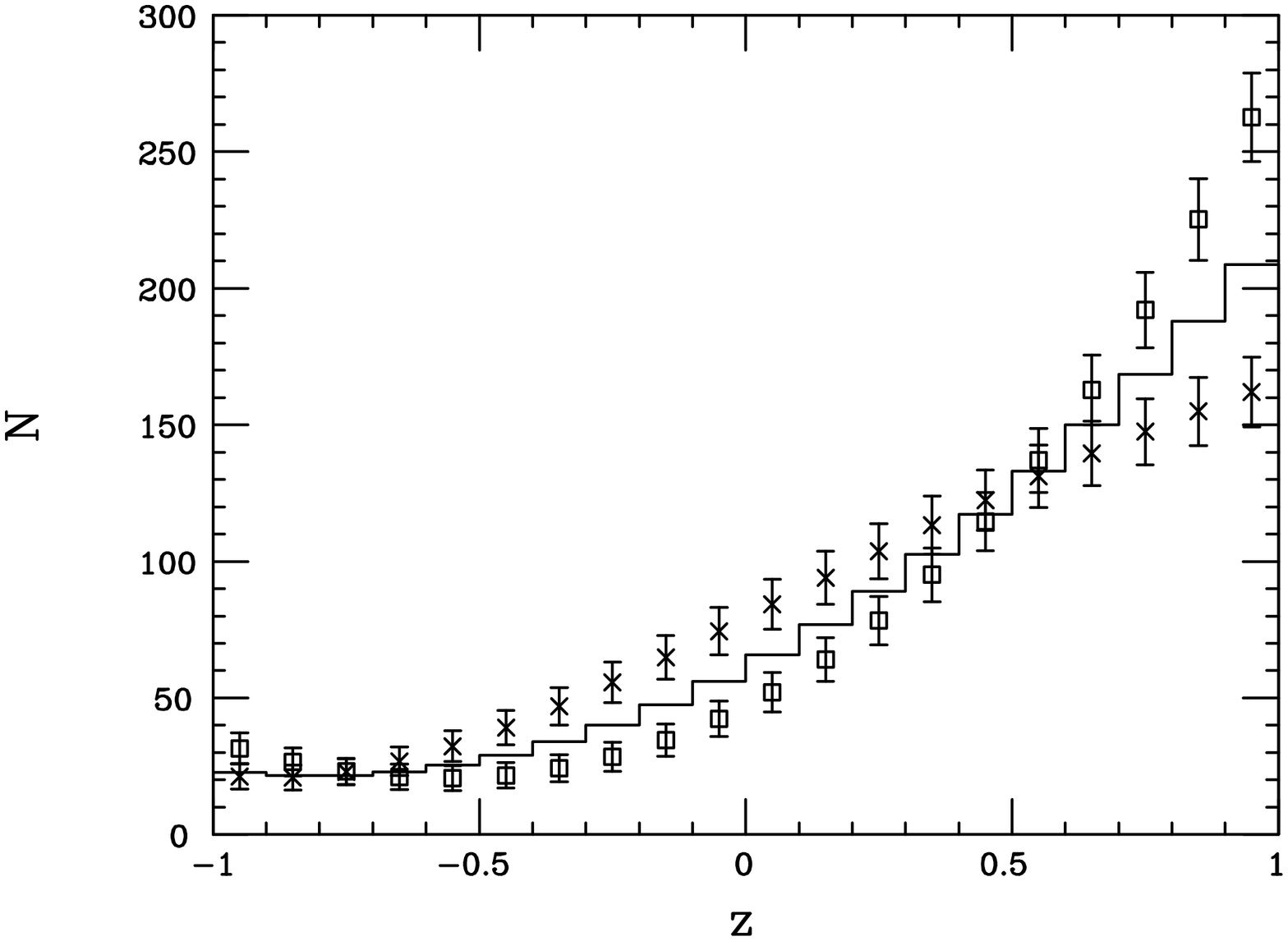,height=6.9cm,width=9cm,angle=-90}}
\vspace*{-.75cm}
\centerline{
\psfig{figure=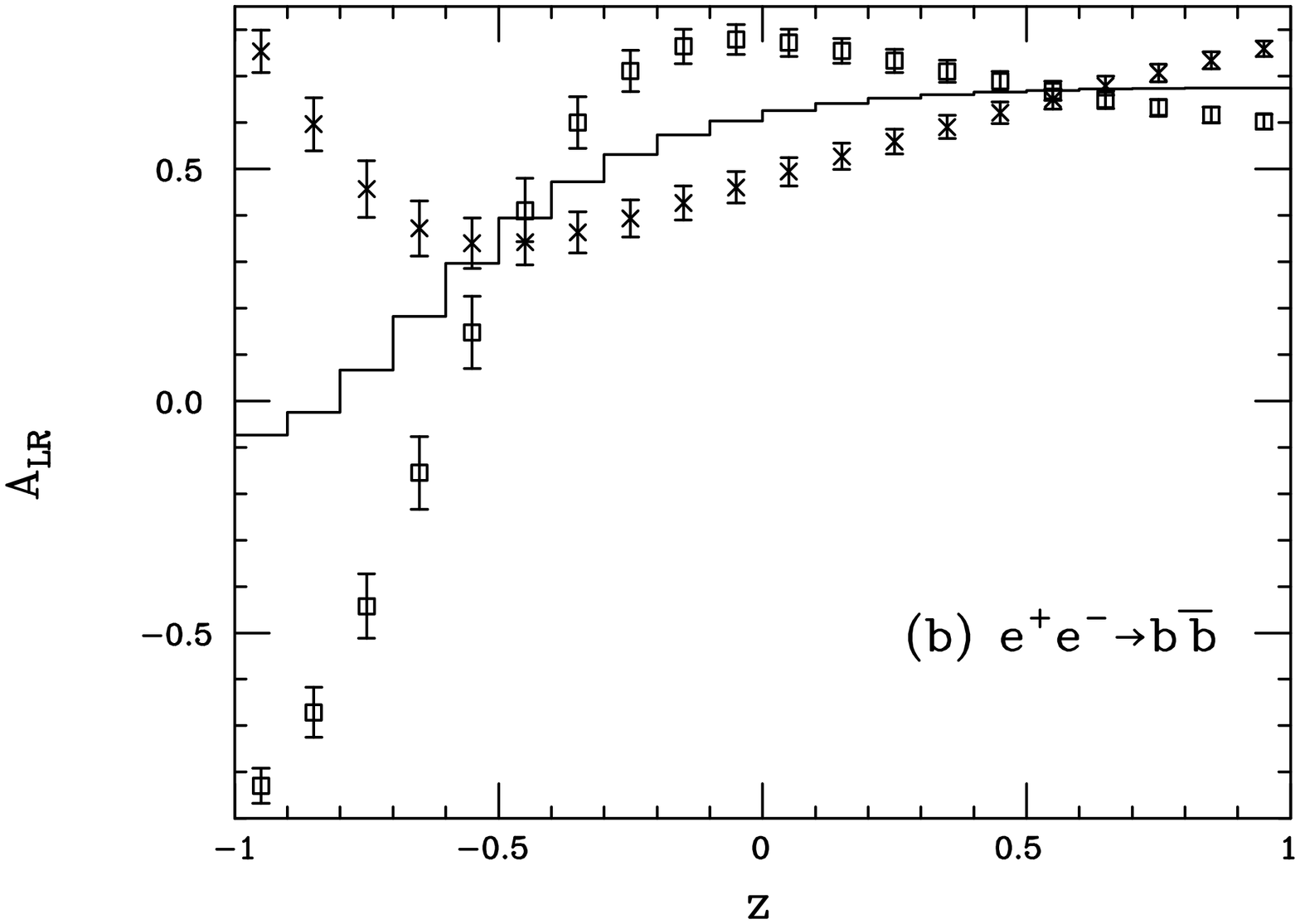,height=6.9cm,width=9cm,angle=-90}}
\vspace*{-0.6cm}
\caption{Distortions(top) in the bin integrated SM cross section(histogram) for 
$e^+e^-\to b\bar b$ at $\sqrt s=200$ GeV with a luminosity of 1 $fb^{-1}$ 
assuming $M_H=0.6$ TeV. The two sets of `data' correspond to $\lambda=\pm 1$.
The corresponding distortions in the $b$ Left-Right Asymmetry at a 500 GeV 
linear collider assuming $M_H=1.5$ TeV are shown in the bottom panel assuming 
a luminosity of 75 $fb^{-1}$.}
\end{figure}

In $e^+e^-$ collisions, as first shown by Hewett~{\cite {pheno1}}, due to the 
spin-2 nature of the gravitons in the tower, angular distributions (and 
polarization asymmetries) become particularly sensitive probes of this 
scenario. For example, the differential cross section for the process 
$e^+e^- \to f\bar f$ now contains both cubic as well as quartic terms in 
$\cos \theta$ and is shown in Fig.1 for $f=b$ at LEP II energies. In all such 
processes the interference between the SM and graviton KK tower exchanges 
is found to vanish when all angles are integrated over thus emphasizing the 
importance of examining differential distributions when trying to constrain 
$M_H$. Hewett also showed that the nature of the spin-2 graviton indirect 
exchange is sufficiently unique as to be easily distinguishable from other 
forms of new physics such as a $Z'$ for values of $M_H$ almost up to the 
search reach.

One can perform a combined fit for $M_H$ by employing the angular 
distributions and polarization asymmetries 
of all kinematically accessible $f\bar f$ final states, as well as the 
polarization of the $\tau$, to obtain a potential search reach for a linear 
collider with the results as shown in Fig.2. 
We see that the reach from such a combined channel analysis of this type is of 
order $M_H \sim 6-7 \sqrt s$. Similar analyses can be performed for both 
Bhabha and Moller scattering but here one finds that systematic errors tend to 
dominate at large luminosities and, with only one channel each, these 
processes do not provide as good a sensitivity to $M_H$ as does the combined 
fit.

\begin{figure}[htbp]
\centerline{
\psfig{figure=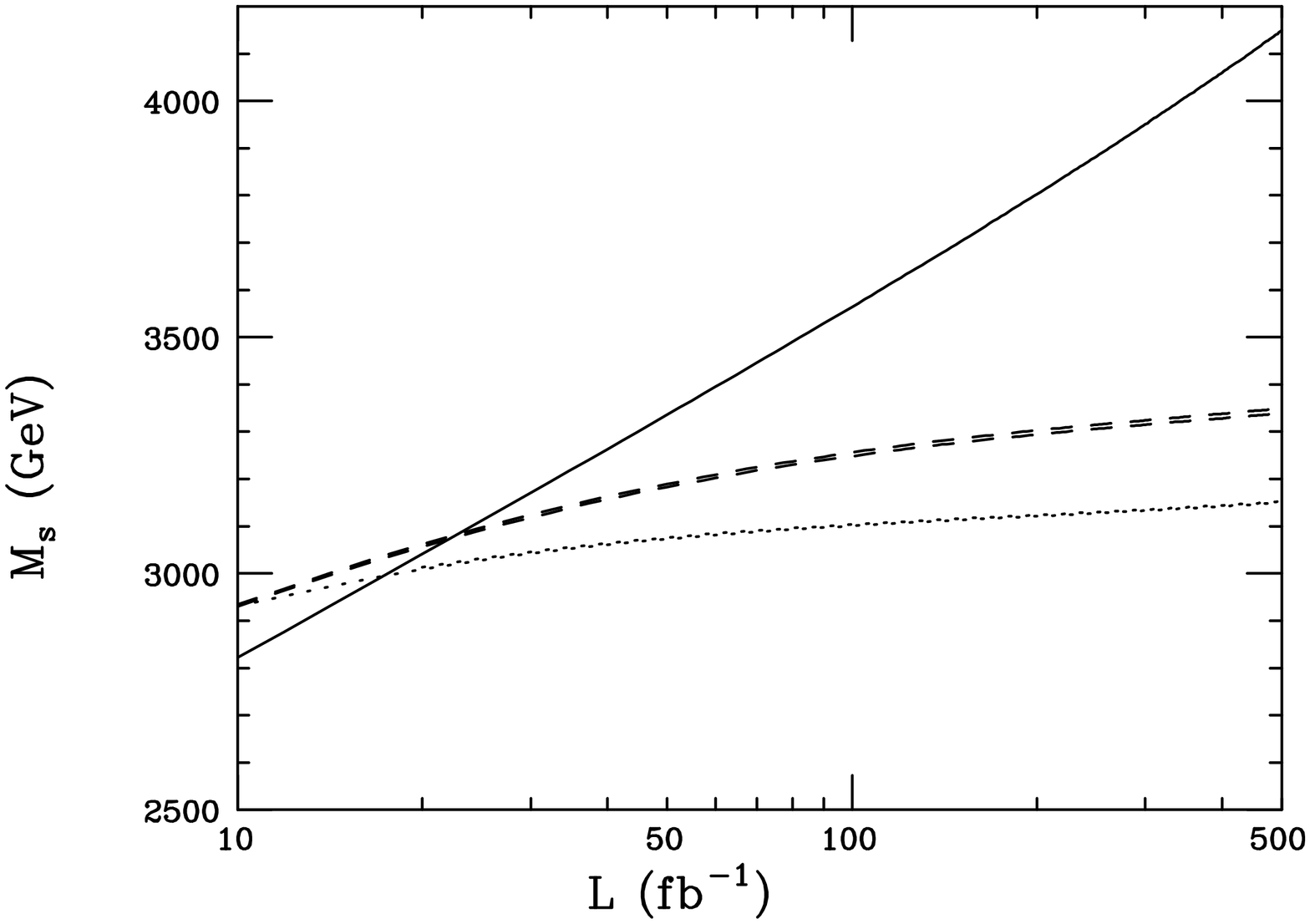,height=6.9cm,width=9cm,angle=-90}}
\vspace*{-.75cm}
\centerline{
\psfig{figure=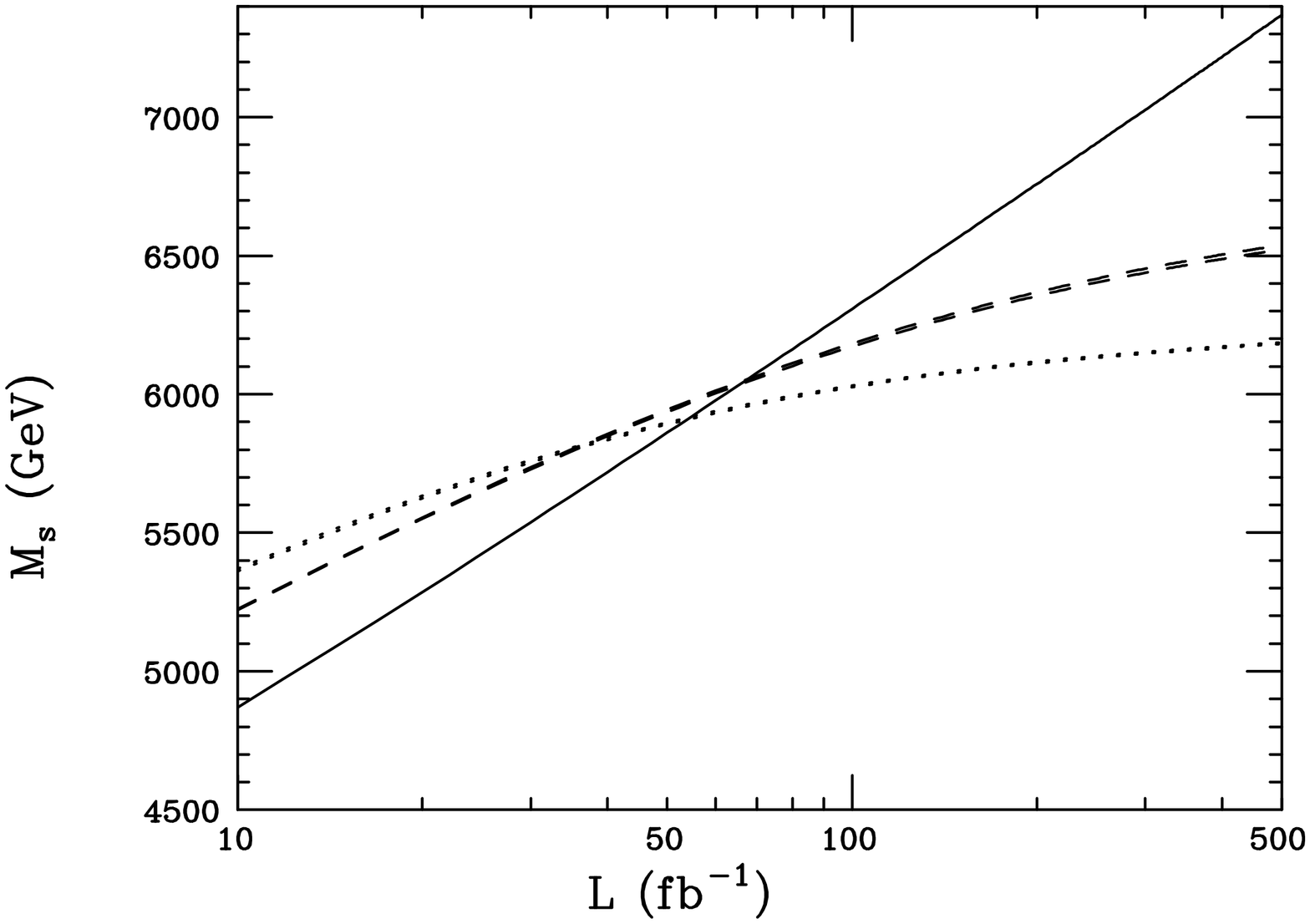,height=6.9cm,width=9cm,angle=-90}}
\vspace*{-0.6cm}
\caption{Search reaches for $M_s(M_H)$ at a 500 GeV(top) and 1000 GeV(bottom) 
$e^+e^-/e^-e^-$ collider as 
a function of the integrated luminosity for Bhabha(dashed) and Moller(dotted) 
scattering for either sign of the parameter $\lambda$ in comparison to the 
`usual' search employing the combined channel 
$e^+e^-\to f\bar f$(solid) fit as described in the text.}
\end{figure}

$e^+e^-$ annihilation into gauge boson pairs also can provide reasonable 
sensitivity to $M_H$ as shown in Fig.3; as can be seen the KK towers do not 
lead to any 
appreciable modification of the $\gamma\gamma,ZZ$ angular distribution or in 
their total cross sections. We estimate that a combined fit to $f\bar f$ and 
gauge boson pair final states may have a reach as high as 
$M_H \sin 9-10\sqrt s$. 

We also note briefly in passing that graviton exchange can lead to new 
processes which are forbidden at the tree level in the SM. A good example of 
this is $e^+e^-\to 2H$ where $H$ is the SM or any one of the neutral MSSM 
Higgs bosons. The rate for such processes has been shown to be potentially 
significant{\cite {pheno1}} at linear colliders.

\begin{figure}[htbp]
\centerline{
\psfig{figure=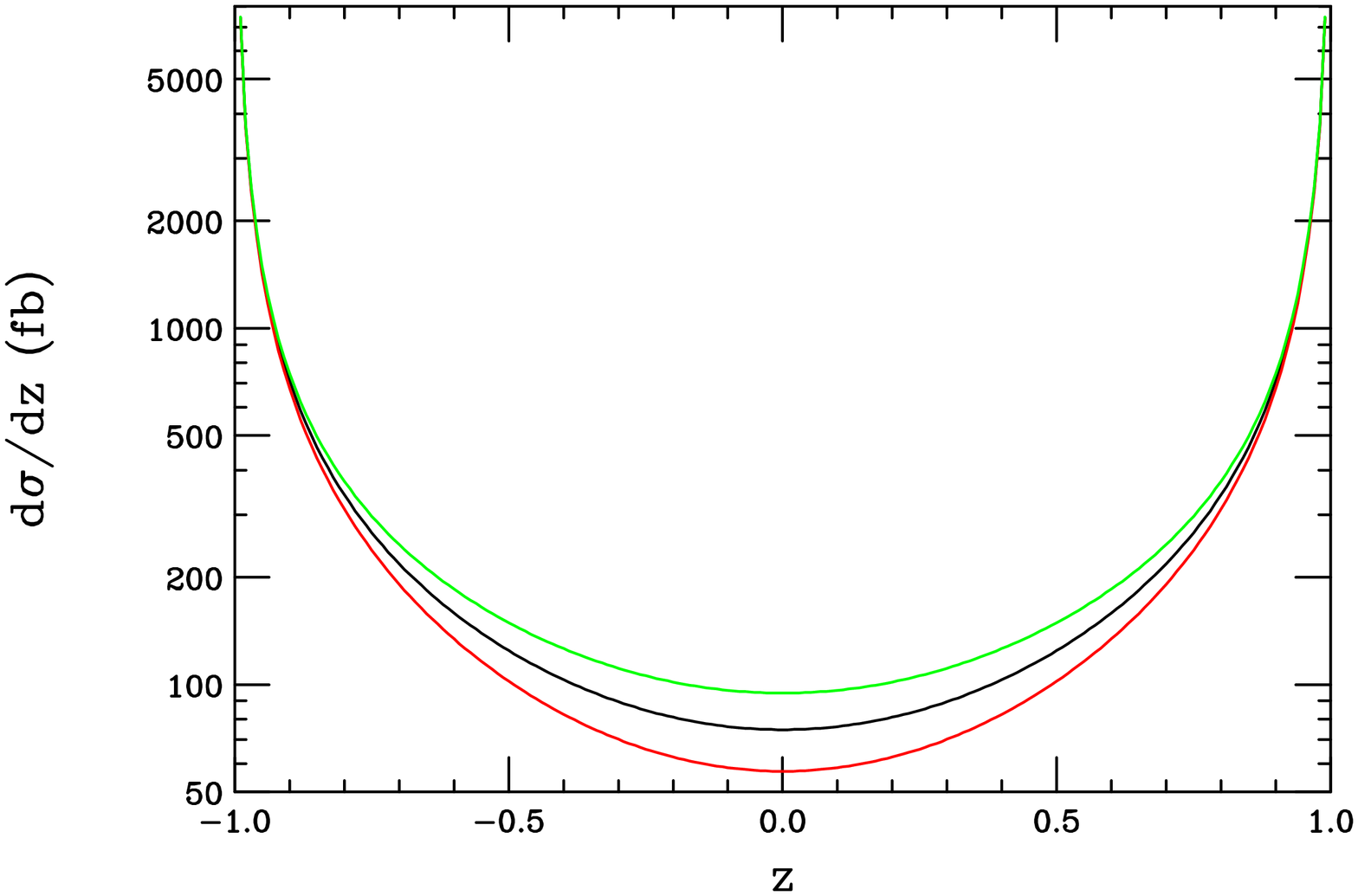,height=6.9cm,width=9cm,angle=90}}
\vspace*{0.1cm}
\centerline{
\psfig{figure=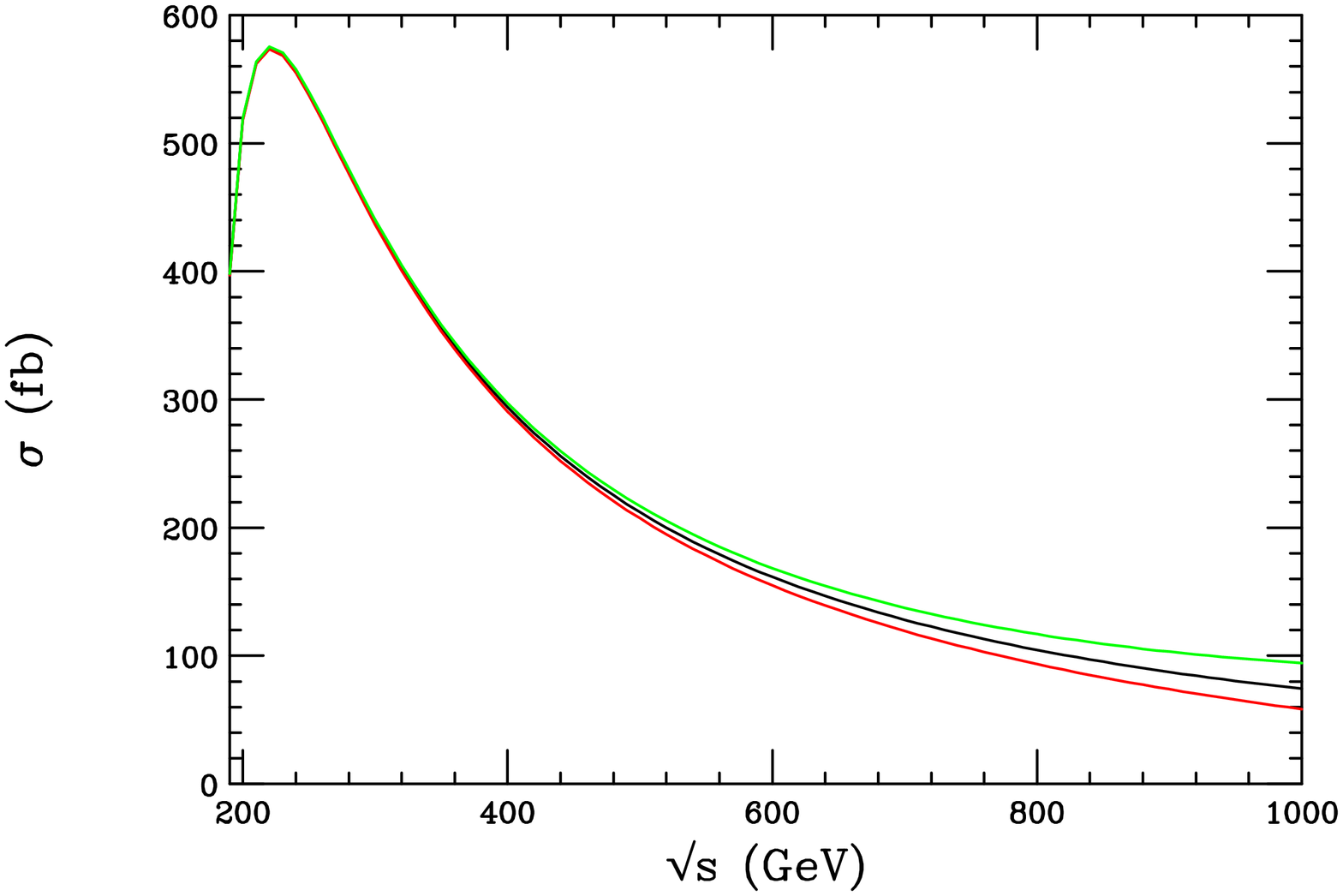,height=6.9cm,width=9cm,angle=90}}
\vspace*{-0.1cm}
\caption{(Top)Differential cross section for $e^+e^-\to \gamma\gamma$ at 1 TeV 
for the SM (center curve) and for $M_H=3$ TeV with $\lambda=\pm 1$. (Bottom) 
$Z$-pair cross section for the same cases as in the Top panel.
Note $z=\cos \theta$.}
\end{figure}

Table 1 provides a useful comparison of the various collider's capabilities 
to probe the value of $M_H$ by employing a number of different processes. Note 
that the process $\gamma \gamma \to W^+W^-$ has the greatest reach of those 
so far examined.

\begin{table}
\centering
\begin{tabular}{|c|c|c|} \hline\hline
Reaction & LEP II (2 fb$^{-1}$) & LC (100 fb$^{-1}$) \\ \hline
$\epem\to f\bar f$ & 1.15 & 6.5$\sqrt s$ \\
$\epem\to\epem$ & 1.0 & 6.2$\sqrt s$ \\
$e^-e^-\to e^-e^-$ &  & 6.0$\sqrt s$ \\
$\epem\to\gamma\gamma$ & 1.4 &  3.2$\sqrt s$ \\
$\epem\to WW/ZZ$ & 0.9 & 5.5$\sqrt s$ \\ \hline
        & Tevatron (2 fb$^{-1}$) & LHC (100 fb$^{-1}$) \\ \hline
$p^(\bar p^)\to \ell^+\ell^-$ & 1.4 & 5.3 \\
$p^(\bar p^)\to t\bar t$ & 1.0 & 6.0 \\
$p^(\bar p^)\to jj$ & 1.0 & 9.0 \\
$p^(\bar p^)\to WW$ & 0.8 & \\
$p^(\bar p^)\to \gamma\gamma$ & 1.4 & 5.4 \\ \hline
       & HERA (250 pb$^{-1}$) & \\ \hline
$ep\to e+$ jet & 1.0  \\ \hline
       & $\gamma\gamma$ Collider (100 fb$^{-1}$) & \\ \hline
$\gamma\gamma\to \ell^+\ell^-/t\bar t/jj$ & 4$\sqrt s$ & \\
$\gamma\gamma\to\gamma\gamma/ZZ$ & $(4-5)\sqrt s$ & \\
$\gamma\gamma\to WW$ & 11$\sqrt s$ & \\ \hline\hline
\end{tabular}
\caption{$M_H$ search limits in TeV for a number of various processes.}
\label{exch}
\end{table}

\section{The RS Model}

Randall and Sundrum(RS){\cite {rs}} have proposed a new 
scenario wherein the hierarchy is generated by an
exponential function of the compactification radius, called a warp factor.
Unlike the ADD model, they assume a 5-dimensional non-factorizable geometry, 
based on a slice
of $AdS_5$ spacetime.  Two 3-branes, one being `visible' with the other being
`hidden', with opposite tensions rigidly reside at
$S_1/Z_2$ orbifold fixed points, taken to be $\phi=0,\pi$, where $\phi$ is
the angular coordinate parameterizing the extra dimension.  It is assumed that 
the extra-dimensional bulk is only populated by gravity and that the SM lies on 
the brane with negative tension. The 
solution to Einstein's equations for this configuration, maintaining
4-dimensional Poincare invariance, is given by the 5-dimensional metric
$ds^2=e^{-2\sigma(\phi)}\eta_{\mu\nu}dx^\mu dx^\nu+r_c^2d\phi^2$, 
where the Greek indices run over ordinary 4-dimensional spacetime, 
$\sigma(\phi)=kr_c|\phi|$ with $r_c$ being the compactification radius of the
extra dimension, and $0\leq |\phi|\leq\pi$.  Here $k$ is a scale of
order the Planck mass and relates the 5-dimensional Planck scale $M$ to the 
cosmological constant. Examination of the action in the 
4-dimensional effective theory in the RS scenario yields the relationship 
$\mpl^2= M^3/k$ for the reduced effective 4-D Planck scale.  

Assuming that we live on the 3-brane located at $|\phi|=\pi$, it is found
that a field on this brane with the fundamental mass
parameter $m_0$ will appear to have the physical mass $m=e^{-kr_c\pi}m_0$.
TeV scales are thus generated from fundamental scales of order $\mpl$
via a geometrical exponential factor and the observed scale hierarchy is
reproduced if $kr_c\simeq 11-12$.  Hence, due to the exponential nature of the
warp factor, no additional large hierarchies are generated. 

Recent analyses{\cite {dhr}} examined the phenomenological 
implications and constraints on the RS model that arise from the exchange of 
weak scale towers of gravitons. There it was shown that the masses of the KK 
graviton states lie at the weak scale are given by $m_n=kx_ne^{-kr_c\pi}$ 
where $x_n$ are the roots 
of $J_1(x_n)=0$, the ordinary Bessel function of order 1. It is important to 
note that these roots are {\it not} 
equally spaced, in contrast to most KK models with one extra dimension, due to 
the non-factorizable metric. This is an important phenomenological distinction. 
Expanding the graviton field into the KK states one finds the interaction 
\begin{equation}
{\cal L} = - {1\over\mpl}T^{\alpha\beta}(x)h^{(0)}_{\alpha\beta}(x)-
{1\over\Lambda_\pi}T^{\alpha\beta}(x)\sum_{n=1}^\infty 
h^{(n)}_{\alpha\beta}(x)\,.
\label{effL}
\end{equation}
Here, $T^{\alpha \beta}$ is the stress energy tensor on the brane and we see 
that the zero mode separates from the sum and couples with the usual
4-dimensional strength, $\mpl^{- 1}$; however, all the 
massive KK states are only suppressed by $\Lambda_\pi^{- 1}$, where we find
that $\Lambda_\pi = e^{- kr_c\pi} \mpl$, which is of order  the weak 
scale. This implies that the tower of weak scale gravitons also couple with 
weak strength and so may have important phenomenological impact. In 
particular, the KK gravitons may now appear as $s-$channel resonances in a 
large number of processes. Unfortunately, no resonances occur in the $e^-e^-$ 
channel; there the physics of the ADD and RS models will appear to be quite 
similar in character. 

The RS model has essentially 2 free parameters which we can take to be the mass 
of the first KK graviton mode and the ratio $c=k/\mpl$; the later quantity is 
restricted to be less than $\sim 0.1$ to maintain the self-consistency of the 
scenario (to prevent a radius of curvature smaller than the Planck scale in 
5 dimensions) and if it is taken too small another hierarchy is formed so that 
very small values must also be avoided. The present and future bounds on the 
parameters of this model can be found in Ref.~{\cite {dhr}} assuming no signal 
is observed. 
Fig. 4 shows the cross section and $A_{FB}$ for the process 
$e^+e^- \to \mu^+\mu^-$ as a function of $\sqrt s$ in the presence of KK 
graviton resonances for several values of the parameter $c$ in the range 
0.01-1. Note that the width of the resonance grows quadratically with the 
value $c$; for fixed $c$ the width increases with the third power of the KK 
resonance mass. Sitting on any one of 
these KK resonances, in the case of small values of $c$, will immediately 
reveal the unique quartic angular distribution corresponding to spin-2 
graviton exchange, \eg, for the case of fermions in the final state one 
obtains a decay distribution 
$\sim 1-3\cos^2 \theta +4\cos^4 \theta$. The branching fractions of these KK 
states are also quite distinctive as is shown in Fig.5.

From the above discussion it is clear that lepton colliders provide an 
excellent means to probe new theories of gravity.

\begin{figure}[htbp]
\centerline{
\psfig{figure=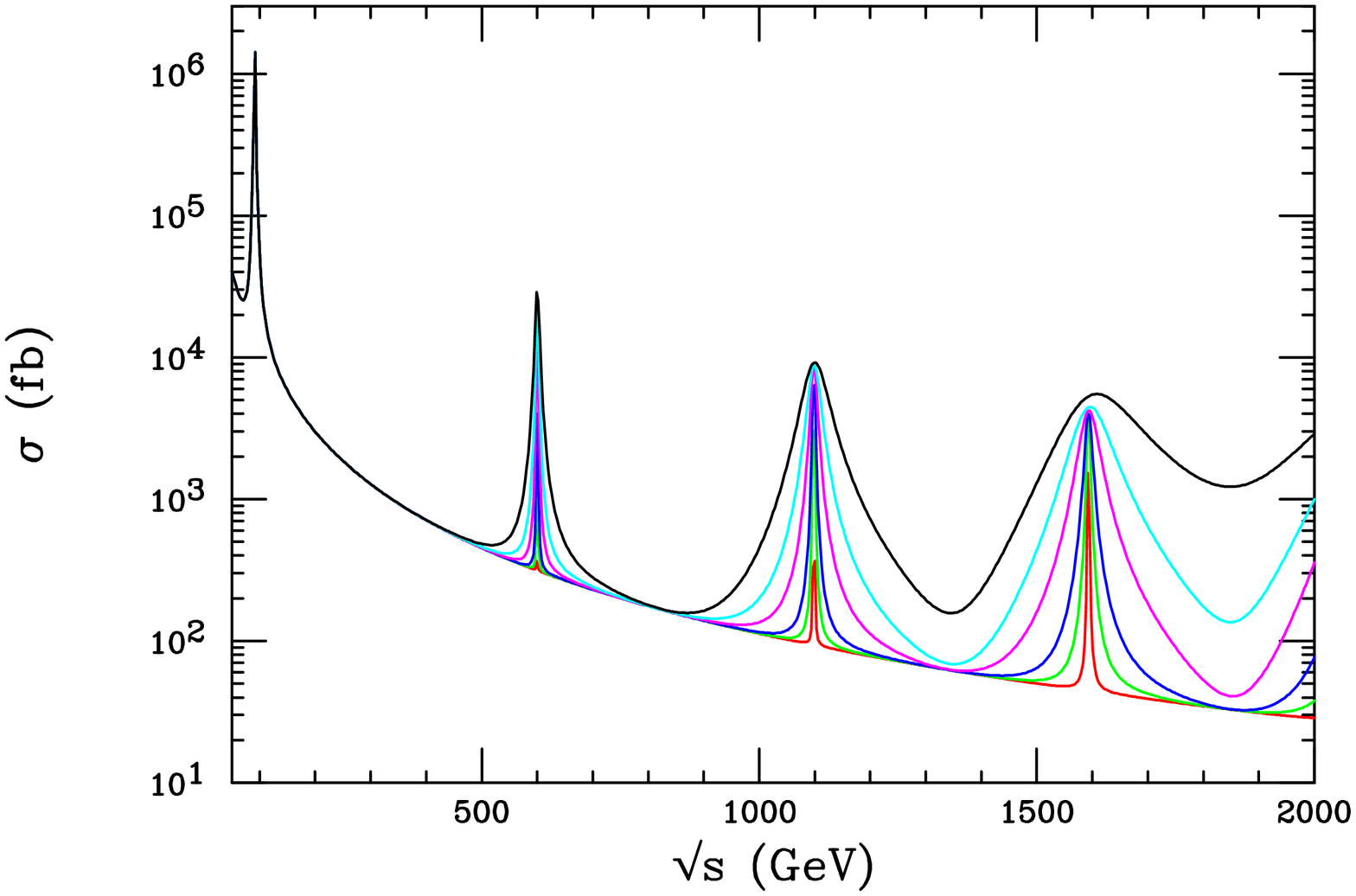,height=6.9cm,width=9cm,angle=90}}
\vspace*{0.1cm}
\centerline{
\psfig{figure=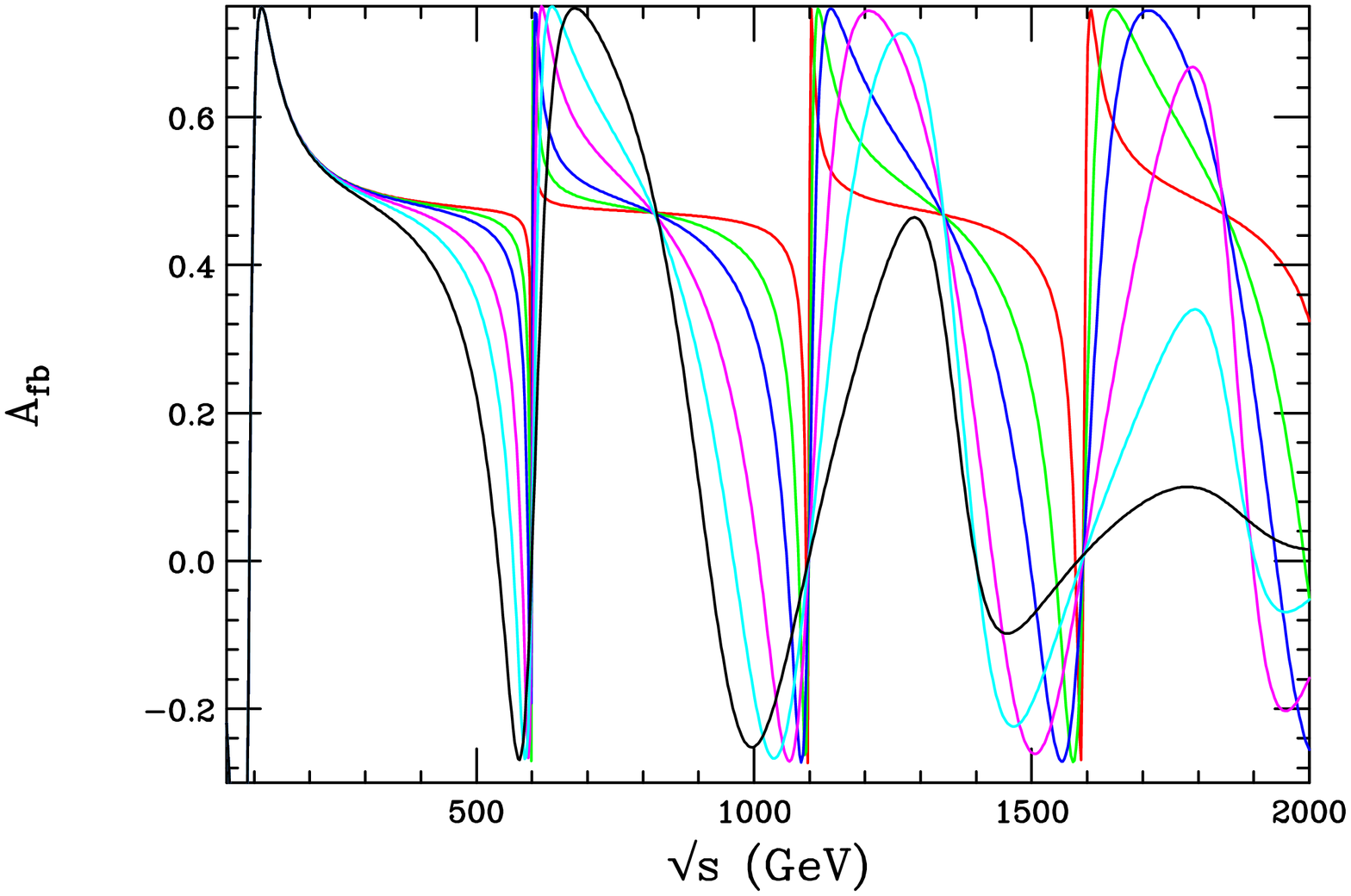,height=6.9cm,width=9cm,angle=90}}
\vspace*{-0.1cm}
\caption{Cross section(top) and $A_{FB}$(bottom) for $e^+e^-\to \mu^+\mu^-$ 
including the exchange of KK gravitons, taking the mass of the first mode to 
be 0.6 TeV, as a function of energy. From outside in the curves correspond to 
c=0.1, 0.07, 0.05, 0.03, 0.02, and 0.01, respectively, with the same labeling 
in the bottom panel.}
\end{figure}

\begin{figure}[htbp]
\centerline{
\psfig{figure=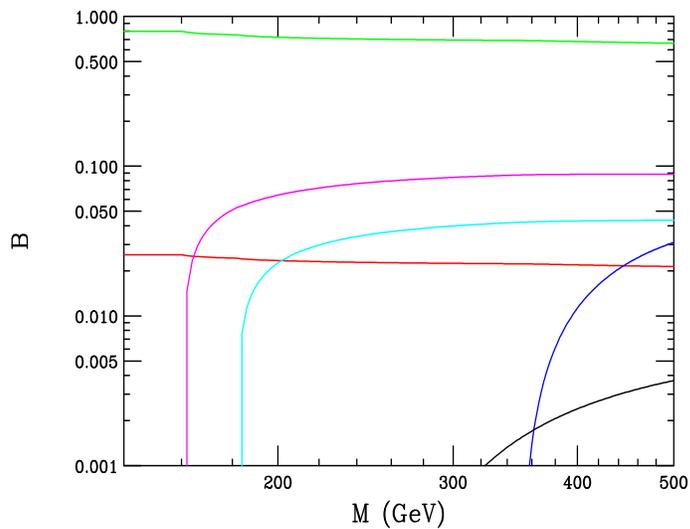,height=6.9cm,width=9cm,angle=90}}
\vspace*{-0.1cm}
\caption{Graviton branching fractions in the RS model as a function of their 
mass. From top to bottom on the right-hand side, the final states are 
jets(gluons and light quarks), W-pairs, Z-pairs, $t\bar t$, lepton pairs and 
Higgs pairs for $M_H$=120 GeV. The branching fraction for photon pairs is 
twice that for leptons.}
\end{figure}

\section*{References}
%
\def\MPL #1 #2 #3 {Mod. Phys. Lett. {\bf#1},\ #2 (#3)}
\def\NPB #1 #2 #3 {Nucl. Phys. {\bf#1},\ #2 (#3)}
\def\PLB #1 #2 #3 {Phys. Lett. {\bf#1},\ #2 (#3)}
\def\PR #1 #2 #3 {Phys. Rep. {\bf#1},\ #2 (#3)}
\def\PRD #1 #2 #3 {Phys. Rev. {\bf#1},\ #2 (#3)}
\def\PRL #1 #2 #3 {Phys. Rev. Lett. {\bf#1},\ #2 (#3)}
\def\RMP #1 #2 #3 {Rev. Mod. Phys. {\bf#1},\ #2 (#3)}
\def\NIM #1 #2 #3 {Nuc. Inst. Meth. {\bf#1},\ #2 (#3)}
\def\ZPC #1 #2 #3 {Z. Phys. {\bf#1},\ #2 (#3)}
\def\EJPC #1 #2 #3 {E. Phys. J. {\bf#1},\ #2 (#3)}
\def\IJMP #1 #2 #3 {Int. J. Mod. Phys. {\bf#1},\ #2 (#3)}

\eject


\begin{thebibliography}{99}
\bibitem{funny}
We denote by $M_H$ the universal scale of the dimension-8 operators which 
control virtual graviton exchange in the ADD model which was introduced by 
Hewett in Ref.~{\cite {pheno1}}.
\end{thebibliography}
\end{document}